\documentclass[11pt]{article} 
\usepackage{moriond,epsfig,axodraw} 

\def\Journal#1#2#3#4{{#1} {\bf #2}, #3 (#4)} 
 
\def\NPB{{\em Nucl. Phys.} B} 
\def\PLB{{\em Phys. Lett.}  B} 
\def\PRL{\em Phys. Rev. Lett.} 
\def\PRD{{\em Phys. Rev.} D}

\def\be{\begin{equation}} 
\def\ee{\end{equation}} 
\def\bea{\begin{eqnarray}} 
\def\eea{\end{eqnarray}} 
 
\def\Rpv{R_p \! \! \! \! \! \! /~~}
\def\beq{\begin{equation}}

\def\eeq{\end{equation}}

\def\bea{\begin{eqnarray}}

\def\eea{\end{eqnarray}}

\def\bq{\begin{quote}}

\def\eq{\end{quote}}

\def\gappeq{\mathrel{\rlap {\raise.5ex\hbox{$>$}}

{\lower.5ex\hbox{$\sim$}}}}

\def\lappeq{\mathrel{\rlap{\raise.5ex\hbox{$<$}}

{\lower.5ex\hbox{$\sim$}}}}

\def\Rp{R_p}

\def\Rpv{R_p \! \! \! \! \! \! /~~}

\def\mnu{[m_{\nu}]_{ij}}

\def\msusy{M_S}

\def\ltap{\raisebox{-.4ex}{\rlap{$\sim$}} \raisebox{.4ex}{$<$}}

\newcommand{\onetwo}{\Delta m^2_{12}}

\newcommand{\onethree}{\Delta m^2_{13}}

\newcommand{\twothree}{\Delta m^2_{23}}

\newcommand{\sun}{\Delta m^2_{\rm solar}}

\newcommand{\atm}{\Delta m^2_{\rm atm}}

\newcommand{\ssqsun}{\sin^2 2\theta_{12}}

\newcommand{\ssqatm}{\sin^2 2\theta_{23}}

\newcommand{\dmui}{\delta^i_{\mu}}

\newcommand{\dmuj}{\delta^j_{\mu}}

\newcommand{\dbi}{\delta^i_{B}}

\newcommand{\dbj}{\delta^j_{B}}

\newcommand{\dlijk}{\delta^{ijk}_{\lambda}}

\newcommand{\dlpipq}{\delta^{ipq}_{\lambda'}}

\def\bea{\begin{eqnarray}}   

\def\eea{\end{eqnarray}}

\begin{document} 
\begin{flushright}
LPT Orsay/03-20\\
SINP/TNP/03-14\\
CI-UAN/03-01T\\
\texttt{hep-ph/0305330}
\end{flushright}

\title{IMPACT OF RECENT NEUTRINO DATA ON R-PARITY VIOLATION} 
 
\author{\sf Asmaa Abada $^{1}$, \underline{\sf Gautam Bhattacharyya}
\footnote{Presented the invited talk. To appear in the proceedings of
38th Rencontres de Moriond (Electroweak Interactions and Unified
Theories), Les Arcs 1800, France, Mar 15-22, 2003.} $^{2}$, \sf Marta
Losada $^{3}$}

\address{$^1$ Laboratoire de Physique Th\'eorique, Universit\'e de
Paris XI, B\^atiment 210, 91405 Orsay Cedex, France \\ $^2$ Saha
Institute of Nuclear Physics, 1/AF Bidhan Nagar, Kolkata 700064, India
\\ $^3$ Centro de Investigaciones, Universidad Antonio Nari\~{n}o,
Cll. 58A No. 37-94, Santa Fe de Bogot\'{a}, Colombia }

\maketitle\abstracts{ We take both the bilinear and trilinear R-parity
violating couplings in supersymmetric models as a source of neutrino
masses and mixings.  Using the solar and atmospheric data and the
Chooz constraint we determine the allowed ranges of those couplings.
We also estimate the effective mass for neutrinoless double beta decay
in this scenario. }

\section{R-parity violation and its parametrization} 
 
In supersymmetric models, lepton and baryon numbers ($L$ and $B$,
respectively) are not automatically conserved, unlike in the standard
model. In terms of $L$ and $B$, one defines R-parity as $\Rp =
(-1)^{3B+L +2S}$, where $S$ is the spin of a particle. Stringent
phenomenological limits on R-parity violating ($\Rpv$)
interactions\cite{rpar1,rpar2} can be found in Ref.~\cite{reviews}. $B$
violation has got nothing to do with neutrino mass generation, and we
assume that such interactions are absent. Neutrino Majorana mass
generation requires two units of $L$ violation. For this purpose, we
allow both bilinear ($\mu_i$) and trilinear ($\lambda, \lambda'$)
interactions in the superpotential as well as the bilinear soft terms
($B_{i}$), given by, \beq W= \mu^J {H}_u L_J + \frac{1}{2}\lambda^{JK
\ell} L_J L_K E^c_{\ell} + \lambda^{'Jpq} L_JQ_p D^c_q + h_u^{pq}
{H}_u Q_p U^c_q \ ,\label{S} \eeq where the vector $L_J = (H_d, L_i)$
with $J:4..1 $.  The soft supersymmetry-breaking potential is
\begin{eqnarray}
V_{soft}& = & \frac{\tilde{m}_u^2}{2} H_u^{\dagger} H_u + \frac{1}{2}
 L^{J \dagger} [\tilde{m}^2_L]_{JK} L^K  + B^J H_u L_J   \nonumber \\ 
& & + A^{ups} H_u Q_p U^c_s + 
     A^{Jps} L_J Q_p D^c_s +
    \frac{1}{2}A^{JKl} L_J L_K E^c_l + {\rm ~h.c.}  \label{soft}
\end{eqnarray}

 It should be noted that field redefinitions of the $H_{d}, L_{i}$
 fields correspond to basis changes in $L_{J}$ space and consequently
 the Lagrangian parameters will be altered.  Hence we use the
 basis-independent parameters constructed in \cite{DL1,DL2} and write
 the neutrino mass matrix in terms of such parameters $\dmui, \dbi,
 \dlijk, \dlpipq$, which in the basis in which the sneutrino vacuum
 expectation values are zero correspond to the Lagrangian parameters
 $\mu^i/|\mu|$, $B^i/|B|$, $\lambda_{ijk}$, $\lambda'_{ipq}$,
 respectively.

This talk is mainly based on the paper \cite{abl}.  For previous
studies of $\Rpv$ effects on neutrino masses, see also
Refs.~\cite{everyone,AM1,AM2}.

\section{Generation of neutrino masses by R-parity violation} 

To understand the effects of the bilinear and trilinear terms on
neutrino mass matrices, let us consider them one by one. First, switch
on only the bilinear $\dmui$ terms which appear in the
superpotential. Such terms generate neutrino masses at the tree level
which are proportional to $\dmui \dmuj$. This constitutes a rank 1
mass matrix which leads to {\em only one} nonzero eigenvalue. This is
not enough since we know that the solar and atmospheric neutrino data
require at least two nonzero eigenvalues. 

Then we turn on the bilinear $\dbi$ terms which appear in the scalar
potential. They contribute to neutrino mass at the loop level via the
Grossman-Haber diagrams \cite{GH}, in which there are gauge couplings
at the neutrino vertices while there are two types of $\Rpv$
interactions giving rise to the $\Delta L = 2$ Majorana mass via the
diagram of Fig.~\ref{fGH}. The first type has $\Rpv$ couplings located
at positions III $+$ IV (slepton-Higgs mixing) with contributions
proportional to $\delta_B \delta_B$. In the second type, the $\Rpv$
interactions are located at positions V $+$ IV (neutrino-neutralino
and slepton-Higgs mixing) with contributions proportional to
$\delta_{\mu} \delta_B$. Now, the $\dmui$ and $\dbi$ terms together
give rise to two nonzero masses, while one eigenvalue still remains
zero. As far as data is concerned, this scenario works \cite{AMS}.
 
Then we turn on the trilinear interaction for $L$ violation, namely,
the $\dlijk$ (or $\dlpipq$) couplings. We then have loops
involving the trilinear $\Rpv$ couplings $\lambda$ or $\lambda'$ at
the neutrino vertices I and II in Fig.~1 (with lepton/slepton or
quark/squark as propagators). They give contributions proportional to
$\delta_{\lambda} \delta_{\lambda}$ (or $\delta_{\lambda'}
\delta_{\lambda'}$).
One also has the diagram of Fig.~\ref{fmulambda}, where two units of $L$
violation come from positions V (neutrino-neutralino mixing) and II
($\lambda$ or $\lambda'$ vertex). Their contribution to the neutrino
mass is proportional to $\delta_{\mu} \delta_{\lambda}$ (or
$\delta_{\mu} \delta_{\lambda'}$). Now, with bilinear and trilinear
terms together, all the neutrinos become massive.

\begin{center}
\begin{figure}[htb]
\unitlength1mm
\SetScale{2.8}
\begin{boldmath}

\begin{center}
\begin{picture}(60,50)(0,-10)
\Line(0,0)(15,0)
\Line(45,0)(15,0)
\Line(60,0)(45,0)
\DashCArc(30,0)(15,0,180){1}

\Text(8,0)[c]{$\bullet$}
\Text(8,3)[c]{$V$}
\Text(15,0)[c]{$\bullet$}
\Text(15,-5)[c]{$I$}
\Text(45,0)[c]{$\bullet$}
\Text(45,-5)[c]{$II$}
\Text(20,11)[c]{$\bullet$}
\Text(16,14)[c]{$III$}
\Text(40,11)[c]{$\bullet$}
\Text(43,14)[c]{$IV$}
\Text(-2,0)[r]{$\nu_i$}
\Text(62,0)[l]{$\nu_j$}
\end{picture}
\end{center}

\end{boldmath}
\caption{ The usual loops ($\Rpv$ at I $+$ II) and the
Grossman-Haber loops ($\Rpv$ at III $+$ IV or V $+$ IV) contributing
to the neutrino mass.  }
\label{fGH}
\end{figure}
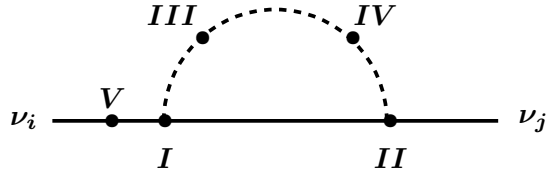

\begin{figure}[htb]
\unitlength1mm
\SetScale{2.8}

\begin{boldmath}
\begin{center}

\begin{picture}(60,60)(0,-20)
\Text(8,0)[c]{$\bullet$}
\Text(8,3)[c]{$V$}
\Text(45,0)[c]{$\bullet$}
\Text(45,-5)[c]{$II$}


\Line(0,0)(15,0)
\Line(45,0)(15,0)
\Line(60,0)(45,0)
\DashCArc(30,0)(15,0,180){1}
\Text(-2,0)[r]{$\nu_i$}
\Text(62,0)[l]{$\nu_j$}
\Text(33,-20)[c]{$v_I$}
\DashLine(33,0)(33,-15){1}
\end{picture}
\end{center}
\end{boldmath}
\caption{Loops with a gauge and a trilinear Yukawa
 coupling. The $\Rpv$ interactions are located at V and II.}
\label{fmulambda}
\end{figure}
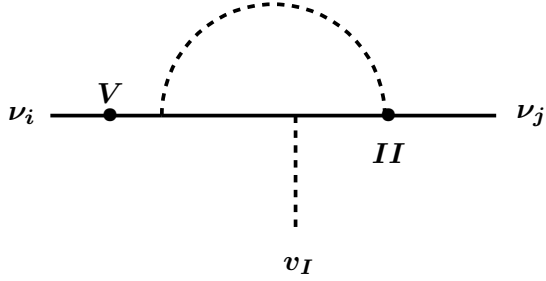

\end{center}

For the sake of simplicity, we set all unknown sparticle masses equal
to $\msusy$ = 100 GeV. We then obtain a neutrino mass matrix of the
form
\bea \mnu &=& \msusy \left[\dmui \dmuj +\frac{\kappa_1}{\cos\beta}
\left(\dmui \dbj + \dmuj \dbi\right) + \frac{\kappa_1}{\cos^2\beta}
\dbi \dbj \right] \nonumber \\ &+& \kappa_2 \left[\sum_{k,n} m_{e_{n}}
m_{e_{k}} \delta_{\lambda}^{ink}\delta_{\lambda}^{jkn} + 3 \sum_{k,n}
m_{d_{n}} m_{d_{k}}
\delta_{\lambda'}^{ink}\delta_{\lambda'}^{jkn}\right] \\ &+& \kappa_3
\left[\sum_{k} m_{e_{k}} (\dmui \delta_{\lambda}^{jkk} + \dmuj
\delta_{\lambda}^{ikk}) + 3\sum_{k} m_{d_{k}} (\dmui
\delta_{\lambda'}^{jkk} + \dmuj \delta_{\lambda'}^{ikk}) \right]
\nonumber, \eea where
\bea \kappa_1 = \frac{g^2}{64\pi^2},
~ \kappa_2 = \frac{1}{8 \pi^2 \msusy},~ \kappa_3 = \frac{g}{16\pi^2\sqrt
2}.  \eea 
We have included for
completeness the contributions arising from the $\delta_{\lambda'}$
terms which we set to zero in our numerical analysis.  The simplest
case to consider with a common $\delta_{\mu}^{i} \equiv
\delta_{\mu},\delta_{B}^{i} \equiv \delta_B $ and $
\delta_{\lambda}^{ink} \equiv \delta_{\lambda}$ does not work as it
cannot accomodate two large mixing angles for
$\theta_{12}$ and $\theta_{23}$.

In our numerical analysis we take : $\delta_{\mu}^{i}$,
$\delta_{B}^{i} $, $\delta_{\lambda}^{ink} \equiv \delta_{\lambda}$,
$\delta_{\lambda'}^{ink}=0$, for $i=$ 1, 2, 3, i.e., seven independent
parameters. Note that a common $\delta_{\lambda}$, in addition to
$\dmui$ and $\dbi$ terms, is enough to make all the neutrinos massive,
and allows us to study how the presence of the trilinear interaction
alters the allowed range of the bilinear parameters.
Thus, the neutrino mass matrix elements will be given by,
\bea m_{11} &=& \msusy\left[(\delta_{\mu}^{1})^{2} +
\frac{\kappa_1}{\cos^2\beta} (\delta_{B}^{1})^{2} +
2\frac{\kappa_1}{\cos\beta}\delta_{\mu}^1 \delta_{B}^1\right]+
2\kappa_3 m_{\tau} \delta_{\mu}^1\delta_{\lambda} +\kappa_{2}
m_{\tau}^{2} \delta_{\lambda}^{2}, \nonumber \\ m_{12} &=&
\msusy\left[\delta_{\mu}^{1}\delta_{\mu}^{2} +
\frac{\kappa_1}{\cos^2\beta} \delta_{B}^{1} \delta_{B}^{2} +
\frac{\kappa_1}{\cos\beta}(\delta_{\mu}^1 \delta_{B}^2 +
\delta_{\mu}^2 \delta_{B}^1)\right]+ \kappa_3 m_{\tau}
(\delta_{\mu}^1\delta_{\lambda} +\delta_{\mu}^2\delta_{\lambda})
+\kappa_{2} m_{\tau}^{2} \delta_{\lambda}^{2}, \nonumber \\ m_{22} &=&
\msusy\left[(\delta_{\mu}^{2})^{2} + \frac{\kappa_1}{\cos^2\beta}(
\delta_{B}^{2})^{2} + 2\frac{\kappa_1}{\cos\beta}\delta_{\mu}^2
\delta_{B}^2\right] + 2\kappa_3 m_{\tau}
\delta_{\mu}^2\delta_{\lambda} +\kappa_{2} m_{\tau}^{2}
\delta_{\lambda}^{2},  \\ 
m_{13} &=&
\msusy\left[\delta_{\mu}^{1}\delta_{\mu}^{3} +
\frac{\kappa_1}{\cos^2\beta} \delta_{B}^{1} \delta_{B}^{3} +
\frac{\kappa_1}{\cos\beta}(\delta_{\mu}^1 \delta_{B}^3 +
\delta_{\mu}^3 \delta_{B}^1)\right] + \kappa_3 m_{\tau}
\delta_{\mu}^3\delta_{\lambda}, \nonumber \\
m_{23} &=&
\msusy\left[\delta_{\mu}^{2}\delta_{\mu}^{3} +
\frac{\kappa_1}{\cos^2\beta} \delta_{B}^{2} \delta_{B}^{3} +
\frac{\kappa_1}{\cos\beta}(\delta_{\mu}^3 \delta_{B}^2 +
\delta_{\mu}^2 \delta_{B}^3)\right] + \kappa_3 m_{\tau}
\delta_{\mu}^3\delta_{\lambda}, \nonumber \\ 
m_{33} &=&
\msusy\left[(\delta_{\mu}^3)^{2} + \frac{\kappa_1}{\cos^2\beta}
(\delta_{B}^3)^{2} + 2\frac{\kappa_1}{\cos\beta}\delta_{\mu}^3
\delta_{B}^3\right], \nonumber\eea 
where we have employed the hierarchy of the
charged fermion masses to keep only the dominant terms, i.e. those
which involve $m_\tau$.

\section{Experimental data on neutrino masses and mixings} 

We write the PMNS matrix, which is 
the rotation matrix from neutrino flavour ($f$) to mass ($i$)
eigenstates, as
 \beq V_{fi} =
\left[
\begin{array}{ccc} c_{12} c_{13} & c_{13} s_{12} &s_{13} \\ -c_{23}
s_{12} - c_{12} s_{13} s_{23} & c_{12} c_{23} - s_{12} s_{13} s_{23} &
c_{13} s_{23} \\ s_{23} s_{12} - c_{12} c_{23} s_{13} & -c_{12} s_{23} -
c_{23} s_{12} s_{13} & c_{13} c_{23}
\end{array} \right],
\label{mixing}
\eeq 
where $c_{ij} \equiv \cos \theta_{ij}$ and $s_{ij} \equiv \sin
\theta_{ij}$. We have neglected the CP phases which are not relevant
for our purpose of extracting allowed ranges of new physics from
oscillation analysis.  We take the solar anomaly to be a consequence
of $\nu_e$-$\nu_\mu$ oscillation, and the relevant mass squared
difference is $\onetwo = \sun$. We consider the atmospheric
oscillation to be between $\nu_\mu$ and $\nu_\tau$, and the relevant
mass squared difference is $\onethree \approx \twothree = \atm$. After
the announcement of the SNO data, the MSW-LMA oscillation is the most
favoured solution with $\sun = (2.5 - 19.0) \times 10^{-5}
~\mathrm{eV}^{2}$ and $\ssqsun = 0.61 - 0.95$. The SuperK atmospheric
neutrino data suggest $\atm = (2 - 5) \times 10^{-3}~\mathrm{eV}^{2}$
with $\ssqatm = 0.88 - 1.0$.  The Chooz and Palo Verde long baseline
reactor experiments constrain $\sin^2 \theta_{13}~\ltap~0.04$.
Tritium $\beta$-decay requires that the absolute mass is
$m_{\nu_e}~\ltap~2.2$ eV. {\sf For the references of all the
experimental data, see \cite{abl}}.

\section{Results and conclusions} 

\begin{figure} 
\hspace{2cm}\epsfxsize=14cm\epsfbox{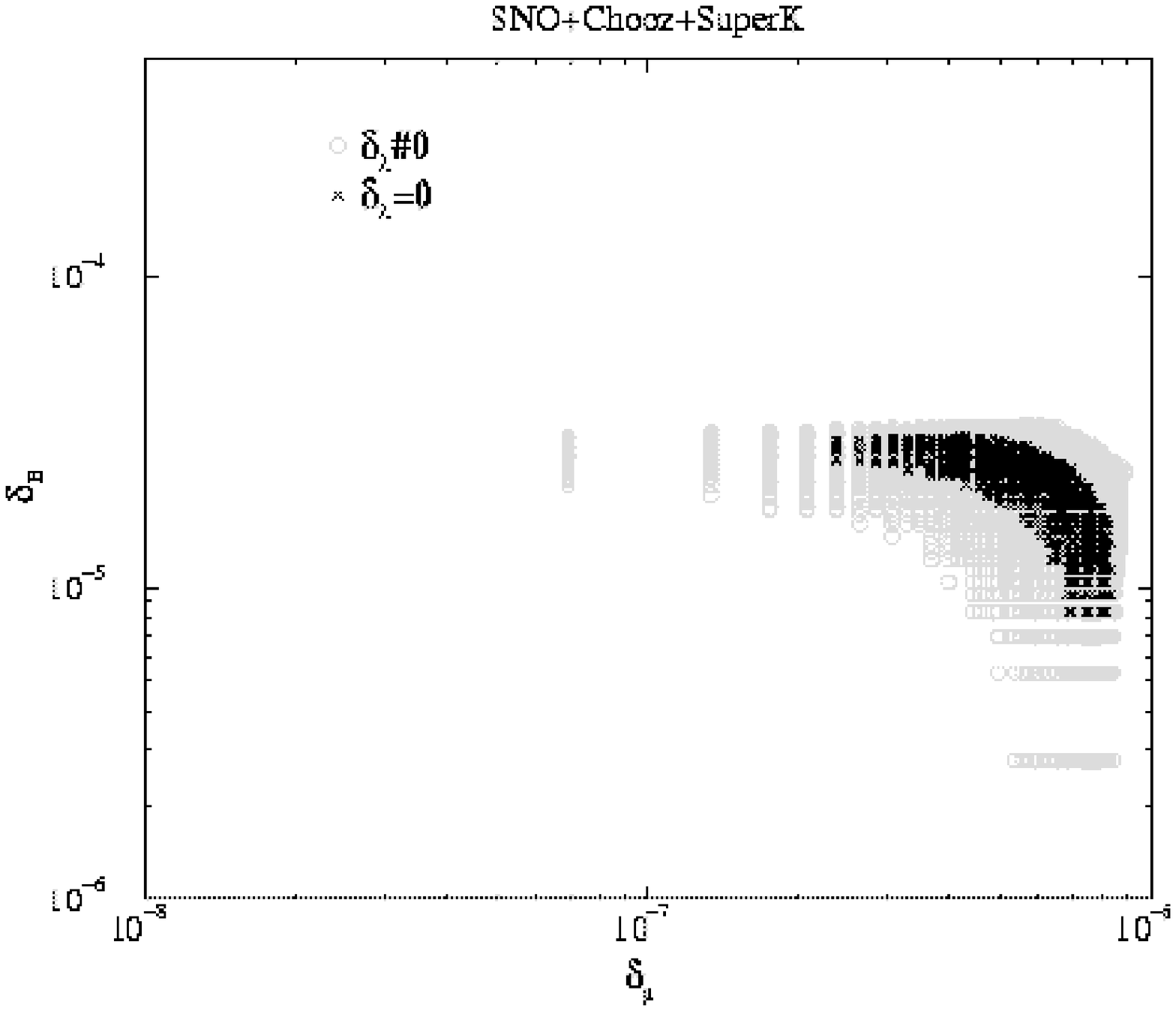}
\caption{We present the allowed region in the $\delta_B \equiv
|\delta_B|$ versus $\delta_{\mu} \equiv |\delta_{\mu}|$ plane. The
circles are solutions for $\delta_{\lambda} \neq 0$, the crosses are
for $\delta_{\lambda}=0$. 
\label{fig:fig3}} 
\end{figure}

\begin{itemize} 
\item 
We have performed a general scan of parameter space made up by the
seven parameters (three $\dmui$, three $\dbi$, one $\delta_{\lambda}$)
that appear in the mass matrix. We allow the tree-level contributions
to either dominate over the loop corrections, to be on the same order
as these, or to be much smaller than the loop terms.  The fitted
values of the couplings, satisfying all the data mentioned in the
previous section, are given in table I.

\item 
In Fig.~\ref{fig:fig3} we have presented the allowed region in the
$|\delta_B| = \sqrt{\sum_i (\dbi)^2}$ versus $ |\delta_{\mu}| =
\sqrt{\sum_i (\dmui)^2}$ plane for the combined fit. We show our
results for both $\delta_{\lambda} \neq 0$ and $\delta_{\lambda}=0$.
It can be seen that the allowed region increases when we admit
non-zero values of $\delta_{\lambda}$. This happens due to the
presence of the $\delta_\mu \delta_{\lambda}$ terms in the mass matrix
(originating from Fig.~\ref{fmulambda}) which can take either sign
thus accomodating a larger region of the parameter space.  This figure
should be compared with Fig.~5 of Ref.~\cite{AMS}. 

\item 
The resulting fit strongly prefers a hierarchical mass pattern in our
scenario, although a distinction between the inverted and the normal
hierarchy is not possible.

\item 
The maximum value of $m_{\rm eff}$ we have predicted (see Table 1) can
hopefully be tested in the next generation of neutrinoless double beta
decay experiments. The range again implies that the spectrum is not
degenerate \cite{ag}.  

\item 
Allowing a non-vanishing $\delta_{\lambda'}$ will not qualitatively
change the pattern of our fit.

\item 
The analysis was done before the first announcement of the KamLAND
results. The pattern of the fit will not be qualitatively altered if
we include the KamLAND data. 

\end{itemize}

\begin{table}[htb]
\begin{center}
\begin{tabular}{|l|l|l|}
\hline 
Couplings & Min & Max\\\hline
$\delta_\lambda $ & $-2.0\times 10^{-4}$  & $2.0\times 10^{-4}$ \\\hline
$\delta_\mu^1$& $-6.8\times 10^{-7}$& $6.8\times 10^{-7}$ \\\hline
$\delta_\mu^2$& $-8.4\times 10^{-7}$& $8.4\times 10^{-7}$\\\hline
$\delta_\mu^3$& $-8.4\times 10^{-7}$& $8.4\times 10^{-7}$ \\\hline
$\delta_B^1$& $-2.7\times 10^{-5}$& $2.7\times 10^{-5}$ \\\hline
$\delta_B^2$& $-3.0\times 10^{-5}$& $3.0\times 10^{-5}$\\\hline
$\delta_B^3$& $-3.0\times 10^{-5}$& $3.0\times 10^{-5}$ \\\hline\hline
$m_{\rm eff} ({\rm eV})$ & $1.9\times 10^{-3}$& $6.7\times 10^{-2}$ \\
\hline
$\sum m_i ({\rm eV}) $ & $4.9\times 10^{-2}$& $0.2$ \\
\hline 
\end{tabular}
\protect\label{seven}
\caption{Allowed range of the couplings satisfying MSW-LMA, SuperK and
Chooz simultaneously (with $\cos\beta=1$).}
\end{center} 
\end{table}

\section*{Acknowledgments} 
I (G.B.) thank the organizers of the Moriond Conference for the
invitation to give this talk, for partial financial assistance, 
as well as for the warm hospitality they provided in such a beautiful 
ambiance in the French Alps!

\newpage 

\end{document}